\documentclass[10pt,journal,english,twocolumn]{IEEEtran}

\usepackage[T1]{fontenc}
\usepackage[utf8]{inputenc}
\usepackage{babel}
\usepackage[babel]{microtype}
\usepackage[babel]{csquotes}

\usepackage{amsmath}
\usepackage{amsfonts}
\usepackage{amssymb}
\usepackage{amsthm}
\usepackage{bm}
\usepackage{mathtools}
\usepackage{dsfont}
\usepackage{siunitx}

\usepackage{graphicx}
\graphicspath{{results/}{figs/}}
\DeclareGraphicsExtensions{.pdf,.png,.jpg}
\usepackage{subcaption}
\usepackage{tikz}
\usepackage{pgfplots}
\pgfplotsset{compat=1.18}

\usepackage[table,svgnames,dvipsnames]{xcolor}

\usepackage{array}
\usepackage{booktabs}
\usepackage{tabularx}
\usepackage{multirow}
\usepackage{pifont}
\newcolumntype{C}{>{\centering\arraybackslash}X}

\usepackage[ruled,vlined]{algorithm2e}

\usepackage[
  backend=biber,
  citestyle=numeric-comp,
  bibstyle=ieee,
  url=false,
  dashed=false,
  isbn=false,
  doi=true
]{biblatex}
\usepackage[acronyms,nomain,xindy,nowarn]{glossaries}
\makeglossaries
\newacronym{5g}{5G}{Fifth-Generation Wireless Systems}
\newacronym{ai}{AI}{artificial intelligence}
\newacronym{ann}{ANN}{artificial neural network}
\newacronym{ask}{ASK}{amplitude-shift keying}
\newacronym[plural={AWGN}, \glsshortpluralkey={AWGN}]{awgn}{AWGN}{additive white Gaussian noise}

\newacronym{bec}{BEC}{binary erasure channel}
\newacronym{ber}{BER}{bit error rate}
\newacronym{bler}{BLER}{block error rate}
\newacronym{bpsk}{BPSK}{binary phase-shift keying}

\newacronym{cdf}{CDF}{cumulative distribution function}
\newacronym{cnn}{CNN}{convolutional neural network}
\newacronym{csi}{CSI}{channel state information}
\newacronym{csit}{CSI\babelhyphen{nobreak}T}{channel-state information at the transmitter}
\newacronym{csir}{CSI\babelhyphen{nobreak}R}{channel-state information at the receiver}

\newacronym{ecc}{ECC}{error-correcting code}
\newacronym{ecdf}{ECDF}{empirical distribution function}
\newacronym{ee}{EE}{energy efficiency}

\newacronym{ff}{FF}{feed-forward}
\newacronym{ffnn}{FF-NN}{feed-forward neural network}
\newacronym{fsk}{FSK}{frequency-shift keying}

\newacronym{gm}{GM}{Gaussian mixture}

\newacronym{iid}{i.i.d.}{independent and identically distributed}
\newacronym{il}{IL}{information leakage}
\newacronym{iot}{IoT}{internet of things}

\newacronym{jcas}{JCAS}{joint communications and sensing}

\newacronym{lb}{LB}{lower bound}
\newacronym{los}{LoS}{line-of-sight}

\newacronym{mac}{MAC}{multiple access channel}
\newacronym{map}{MAP}{maximum a posteriori}
\newacronym{mc}{MC}{Monte Carlo}
\newacronym{mimo}{MIMO}{multiple-input multiple-output}
\newacronym{miso}{MISO}{multiple-input single-output}
\newacronym{ml}{ML}{machine learning}
\newacronym{mmwave}{mmWave}{millimeter wave}
\newacronym{mop}{MOP}{multi-objective programming problem}
\newacronym{mrc}{MRC}{maximum ratio combining}
\newacronym{msc}{MSC}{modulation and coding scheme}
\newacronym{mse}{MSE}{mean squared error}

\newacronym{nlos}{NLoS}{non-line-of-sight}
\newacronym{nn}{NN}{neural network}
\newacronym{nve}{NVE}{normalized validation error}

\newacronym{pdf}{PDF}{probability density function}
\newacronym{pwc}{PWC}{polar wiretap code}

\newacronym{qam}{QAM}{quadrature amplitude modulation}

\newacronym{ra}{RA}{rearrangement algorithm}
\newacronym{rbf}{RBF}{radial basis function}
\newacronym{relu}{ReLU}{rectified linear unit}
\newacronym{ris}{RIS}{reconfigurable intelligent surface}
\newacronym{rnn}{RNN}{recurrent neural network}

\newacronym{sgd}{SGD}{stochastic gradient descent}
\newacronym{sgp}{SGP}{secure goodput}
\newacronym{simo}{SIMO}{single-input multiple-output}
\newacronym{sinr}{SINR}{signal-to-interference-plus-noise ratio}
\newacronym{siso}{SISO}{single-input single-output}
\newacronym{snr}{SNR}{signal-to-noise ratio}
\newacronym{svm}{SVM}{support vector machine}

\newacronym{tvd}{TVD}{total variation distance}

\newacronym{uav}{UAV}{unmanned aerial vehicle}
\newacronym{ub}{UB}{upper bound}
\newacronym{urllc}{URLLC}{ultra-reliable low latency communication}

\newacronym{v2v}{V2V}{vehicle-to-vehicle}
\newacronym{v2x}{V2X}{vehicle-to-everything}

\newacronym{zoc}{ZOC}{zero-outage capacity}
\setacronymstyle{long-short}
\glsdisablehyper

\DeclareUnicodeCharacter{03B5}{\ensuremath{\epsilon}}

\title{Incremental Learning-Based Open-Set Classification of Unknown UAVs via RF Signal Semantics%
\thanks{This work was supported in part by the Vinnova-funded project SLDS: Self-Learning Drone Surveillance, and in part by the CELTIC-NEXT project 3D-NET (3D Networks for 6G Mobile Communications Applications) and Robust and AI-Native 6G for Green Networks (RAI-6Green), with funding provided by Vinnova, the Swedish Innovation Agency.}
}

\author{%
Julie Liu$^{1}$,
Irshad A. Meer$^{1}$,
Cicek Cavdar$^{1}$,
and Mustafa Ozger$^{2}$\\[1mm]
$^{1}$Department of Computer Science, KTH Royal Institute of Technology, Sweden\\
$^{2}$Department of Electronic Systems, Aalborg University, Denmark\\
Email: \{xueliu, iameer, cavdar\}@kth.se, mozger@es.aau.dk
}

\begin{document}
\maketitle
\thispagestyle{empty}
\pagestyle{empty}
\begin{abstract}\noindent\boldmath
The proliferation of civilian and commercial unmanned aerial vehicles (UAVs) has heightened the demand for reliable radio frequency (RF)-based drone identification systems that can operate under dynamic and uncertain airspace conditions. 
Most existing RF-based recognition methods adopt a closed-set assumption, where all UAV types are known during training. 
Such an assumption becomes unrealistic in practical deployments, as new or unknown UAVs frequently emerge, leading to overconfident misclassifications and inefficient retraining cycles. 
To address these challenges, this paper proposes a unified incremental open-set learning framework for RF-based UAV recognition that enables both novel class discovery and incremental adaptation. 
The framework first performs open-set recognition to separate unknown signals from known classes in the semantic feature space, followed by an unsupervised clustering module that discovers new UAV categories by selecting between K-Means and Gaussian Mixture Models (GMM) based on composite validity scores. 
Subsequently, a lightweight incremental learning module integrates the newly discovered classes through a memory-bounded replay mechanism that mitigates catastrophic forgetting. 
Experiments on a real-world UAV RF dataset comprising 24 classes (18 known and 6 unknown) show effective open-set detection, promising clustering performance under the evaluated noise settings, and stable incremental adaptation with minimal storage cost, supporting the potential of the proposed framework for open-world UAV recognition.
\end{abstract}

\glsresetall

\section{Introduction}\label{sec:introduction}
Unmanned aerial vehicles (UAVs) are increasingly used in civilian, commercial, and defense applications, raising concerns about privacy, security, and airspace safety~\cite{10398469,8904324}. This motivates reliable RF-based identification systems that can recognize both known and previously unseen drones.
Among existing sensing modalities, RF-based drone recognition is attractive because it operates passively and is less dependent on lighting or visibility conditions~\cite{svanstrom2021real,Bernardini2017_AcousticSignature,Matson2019_MultiAcousticNodes,Jeon2017_DroneSoundDNN,guvenc2018detection,andravsi2017night}. Recent learning-based RF methods achieve high accuracy in closed-set settings, but typically assume that all UAV types are known during training~\cite{medaiyese2021machine,inani2023machine}.

In real-world airspace, the environment is inherently open, with new or modified UAV types continually emerging due to hardware, firmware, or model updates.
Traditional classifiers trained under the closed-set assumption, where all test samples belong to one of the known training classes, are not suited for such conditions. 
When presented with previously unseen signals, they often produce overconfident predictions for incorrect classes, compromising safety and reliability \cite{geng2020recent}. 
In contrast, open-set recognition (OSR) frameworks relax this assumption by allowing test samples to originate from unknown categories \cite{fu2025reason, yu2024open}. 
An OSR model must not only classify samples from known classes accurately but also detect unseen inputs and flag them as unknown. 
However, most existing OSR approaches in both vision and RF domains focus solely on the rejection of unknown samples, that is, they only detect unfamiliar inputs but do not further characterize or learn from them \cite{yu2024open}.

A natural extension is incremental learning (IL), which incorporates newly discovered classes while preserving previously learned knowledge~\cite{van2022three}. However, stable IL for RF-based UAV recognition remains challenging due to signal variability, environmental dependence, and catastrophic forgetting.
IL for RF-based drone recognition poses unique challenges, as the data are complex-valued, environment-dependent, and affected by hardware and propagation variability. Although IL offers a mechanism for adapting RF classifiers to newly emerging classes, the stochastic and non-stationary nature of wireless signals makes stable incremental adaptation particularly challenging.
While OSR introduces the clustering-based new class discovery to partition detected unknown samples into multiple groups, each comprising signals with similar characteristics \cite{swinney2022k, hoang2019detection, park2024k}. However, most existing studies treat clustering as a standalone process rather than as part of an integrated learning pipeline that connects OSR and IL.
Moreover, in practical RF environments, the pool of unknown samples available for novel class discovery is often limited and noisy, demanding clustering algorithms that are effective and computationally lightweight.
Finally, in field-deployable systems, memory and latency constraints prohibit large-scale retraining or exhaustive replay, demanding mechanisms for selective rehearsal.

To address these challenges, this paper proposes a unified incremental open-set learning framework for RF-based UAV recognition. 
The proposed approach bridges open-set recognition, clustering-based novel class discovery, and replay based incremental learning within a single, scalable pipeline. 
Specifically, an open-set recognition module first separates unknown UAV signals from known classes in a learned semantic feature space using dilated convolution encoders and transformer encoders. 
The rejected samples are then grouped into potential new UAV categories through an unsupervised clustering stage that evaluates K-Means and Gaussian Mixture Models (GMM) as candidate models and selects the result with the best composite validity score.
Finally, a lightweight IL module updates the classifier to incorporate these new classes while retaining performance on known categories through a compact replay strategy that mitigates catastrophic forgetting. 

The main contributions of this paper are summarized as follows:
\begin{itemize}
    \item A unified learning framework for RF-based UAV recognition that integrates open-set detection, clustering-based novel class discovery, and incremental learning for adaptation to previously unseen UAV types.
    \item A robust clustering strategy with composite validity metrics and automatic model-order selection to ensure reliable novel class discovery under noisy RF conditions.
    \item A memory-efficient incremental learning mechanism with bounded replay that improves knowledge retention during adaptation and mitigates catastrophic forgetting.
\end{itemize}

\section{System Model and Problem Formulation}\label{sec:system}
We consider an open-world RF-based UAV recognition scenario, where a spectrum monitoring system continuously observes signals originating from both known and previously unseen drones.
The received RF signal samples are defined as:
\begin{equation}
\label{eq:rf_signal}
x(t)=s_c(t)e^{j2\pi f_c t}+n(t),
\end{equation}
where $s_c(t)$ denotes the transmitted waveform at carrier frequency $f_c$, and $n(t)$ represents additive noise.
These raw in-phase and quadrature (I/Q) signals are first transformed into time–frequency representations through the short-time Fourier transform (STFT), yielding a spectrogram $X(n,w)$ that captures both temporal evolution and spectral structure. 
A logarithmic power scaling and normalization to $[0,1]$ are applied to ensure consistent amplitude ranges across different UAVs and flight conditions.

Let $\mathcal{C}_0$ denote the set of known UAV classes used for training, and let $\mathcal{C}_u$ denote the unknown UAV classes that may appear during deployment, with $\mathcal{C}_u \cap \mathcal{C}_0 = \emptyset$ and $\mathcal{C}_u \cup \mathcal{C}_0 = \mathcal{C}_t$. 
To mitigate catastrophic forgetting, a lightweight replay memory $\mathcal{M}$ is maintained, which stores a limited number of representative samples from previously learned classes under the storage constraint $|\mathcal{M}| \leq M_{\max}$, where $M_{\max}$ denotes the maximum number of representative samples that can be retained from earlier classes.

The objective is to design a learning model that can (i) classify UAVs belonging to known categories, (ii) detect and isolate unknown UAV signals, (iii) cluster the unknown signals into potential new categories, and (iv) incrementally incorporate the newly discovered categories without catastrophic forgetting. 

\subsection{Problem Definition}
Formally, given a stream of UAV RF signals $\{x_t\}$ arriving over time with dynamically evolving class sets $\mathcal{C}_t$, the objective is to learn a classifier $h_{\theta}$ with model parameters $\theta$ that maximizes the recognition performance over both old and newly discovered classes while operating under bounded storage and computational resources:
\begin{equation}
\label{eq:opt_prob}
\begin{aligned}
\max_{\theta} \; & \Big[\underbrace{\mathrm{Acc}_{\text{old}}(\theta)}_{\text{Known classes accuracy}}
+ \underbrace{\mathrm{Acc}_{\text{new}}(\theta)}_{\text{New classes accuracy}} \Big] \\
\text{s.t.}\quad 
& |\mathcal{M}| \le M_{\max}, \\
& \Psi_{\text{comp}}(\theta) \le \Gamma_{\max}.
\end{aligned}
\end{equation}
where $\mathrm{Acc}_{\text{old}}$ and $\mathrm{Acc}_{\text{new}}$ denote the classification accuracies on previously known and newly discovered classes, respectively. $\Psi_{\text{comp}}(\theta)$ denotes the computational cost of the model, such as incremental update overhead or inference complexity, and $\Gamma_{\max}$ represents the maximum allowable compute budget. This formulation captures the joint requirements of recognition, discovery, and incremental adaptation under bounded memory and computation, and provides the basis for the methodology presented next.

\section{Proposed Methodology}\label{sec:method}

\begin{figure*}[t]
  \centering
    \includegraphics[width=\linewidth]{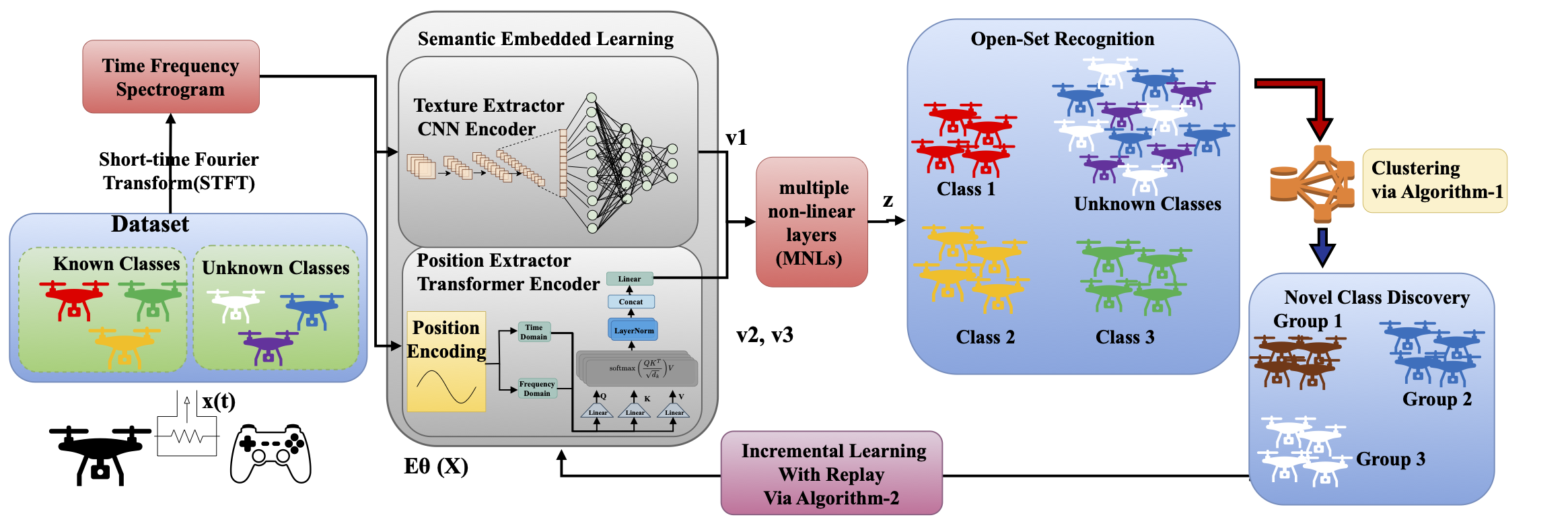}
    \caption{An overview of the proposed open-world RF-based UAV recognition framework integrating open-set recognition, novel class discovery, and incremental learning with replay.}
    \label{fig:flowchart}
\end{figure*}

We first transform the spectrogram $X(n,w)$ into semantic feature representations and then process it by the OSR module, which separates known and unknown classes. The detected unknown samples are subsequently passed to a clustering-based new class discovery module, where they are grouped according to their feature similarity. These clustered samples, together with the existing knowledge base, are then fed into an IL block that updates the main classifier to recognize newly discovered classes without retraining from scratch.
Figure~\ref{fig:flowchart} illustrates the overall architecture of the proposed incremental open-set learning framework.
In the following, we expand in detail on all the components and provide the proposed algorithms.

\subsection{Semantic Embedding Learning}
The first stage aims to learn a discriminative embedding that captures the semantic structure of UAV RF signals. 
Each normalized spectrogram $X(n,w)$ is processed by a hybrid encoder $E_{\theta}(\cdot)$ parameterized by $\theta$, 
which follows a dual-branch architecture comprising a convolutional texture extractor and a transformer-based positional encoder. 
The texture extractor (CNN encoder) produces a texture feature vector $v_1$ through multiscale dilated convolutions that capture fine-grained spectral textures and wideband temporal envelopes. 
In parallel, the transformer encoder generates position-aware representations $v_2$ and $v_3$ from successive self-attention layers using sinusoidal positional encodings to preserve time–frequency locality and model long-range dependencies. 
The intermediate features $\{v_1, v_2, v_3\}$ are then fused and projected through multiple non-linear layers (MNLs) to form a compact semantic embedding \cite{yu2024open}:
\[
\mathbf{z} = f_{\text{MNL}}\big([v_1, v_2, v_3]\big) = E_{\theta}(X) \in \mathbb{R}^D,
\]
where $D$ denotes the dimensionality of the learned embedding space. 
This fused representation $\mathbf{z}$ jointly encodes spectral–temporal textures, positional dependencies, and semantic correlations of the input RF signal.
To ensure that samples of the same UAV type form compact clusters and those from different UAVs are well separated, 
the encoder is trained using a composite feature loss, which is calculated as follows \cite{yu2024open}:
\begin{equation}
L_{\text{feat}} = \eta_1 L_{\text{cen}} + \eta_2 L_{\text{sep}} + \eta_3 L_{\text{CE}},
\label{eq:feature_loss}
\end{equation}
where $L_{\text{cen}}$ enforces intra-class compactness, $L_{\text{sep}}$ imposes inter-class margin separation, 
and $L_{\text{CE}}$ denotes standard cross-entropy supervision. 
The coefficients $\eta_1$, $\eta_2$, and $\eta_3$ control the relative contribution of each component. The model parameters are updated by minimizing a weighted sum of all three losses using stochastic gradient descent (SGD). This optimization yields a well-structured semantic embedding space in which UAVs with similar signal characteristics cluster closely together, 
providing a robust foundation for subsequent open-set recognition and incremental learning.

\subsection{Open-Set Recognition}
After training on the known UAV classes $\mathcal{C}_0$, the learned semantic embeddings are modeled using Gaussian class statistics. 
Each class $k$ is represented by its mean vector $\boldsymbol{\mu}_k$ and covariance matrix $\boldsymbol{\Sigma}_k$, which jointly define a probabilistic decision region in the embedding space. 
During inference, a test embedding $\mathbf{z}$ is compared against all known classes using the Mahalanobis distance, which is calculated as follows \cite{yu2024open}:
\begin{equation}
d_k(\mathbf{z}) = \sqrt{(\mathbf{z}-\boldsymbol{\mu}_k)^{\top}\boldsymbol{\Sigma}_k^{-1}(\mathbf{z}-\boldsymbol{\mu}_k)}.
\label{eq:6}
\end{equation}

For each class, a distance threshold $\tau_k$ is derived using the classical three-sigma rule, assuming that intra-class Mahalanobis distances approximately follow a normal distribution \cite{pukelsheim1994three}, \cite{zhao2013outlier}:
\begin{equation}
\tau_k = \mu_{d_k} + 3\sigma_{d_k},
\end{equation}
where $\mu_{d_k}$ and $\sigma_{d_k}$ denote the mean and standard deviation of the distances within class $k$. 
This three-sigma rule defines a $99.7\%$ confidence boundary, allowing each class to self-adaptively determine its acceptance region in the semantic space. 

At test time, the sample is assigned to the class 
\[
k^* = \arg\min_k d_k(\mathbf{z})
\]
if $d_{k^*}(\mathbf{z}) < \tau_{k^*}$; otherwise, it is rejected as an unknown UAV and added to the candidate pool $\mathcal{U}$ for further clustering. 
This formulation provides a statistically grounded and parameter-free mechanism for open-set recognition, ensuring that samples dissimilar to all known classes are automatically filtered for novel class discovery.

\subsection{Clustering for Novel Class Discovery}
After OSR, embeddings of rejected RF samples are grouped using unsupervised clustering to identify potential new UAV classes. As shown in Algorithm~\ref{alg:cluster_unknowns}, prior to clustering, the feature matrix $Z$ is standardized, and dimensionality reduction is applied when necessary. 
Specifically, Principal Component Analysis (PCA) is employed if the original embedding dimension $d$ exceeds 64, projecting the data to $m = \min(64, d)$ components.  PCA compresses the high-dimensional semantic embeddings while retaining most of their variance, which improves clustering stability and computational efficiency by mitigating the curse of dimensionality. We employ K-Means and Gaussian Mixture Models (GMM) to cluster the rejected unknown UAV samples.
The optimal cluster number $k^{\star}$ is determined through two complementary criteria: the elbow method and a composite validity score
The elbow method detects the inflection point of the inertia curve to distinguish between under and over segmentation \cite{bholowalia2014ebk}.  
The composite score $Q(k)$ integrates multiple clustering quality indices like Silhouette ($S_k$), Calinski–Harabasz ($CH_k$), Davies–Bouldin ($DB_k$), and explained variance ($V_k$). The final composite score $Q(k)$ is defined as:
\begin{equation}
\label{eq:composite_score}
   Q(k)=0.4S_k+0.3\frac{CH_k}{1000}+0.2\frac{1}{1+DB_k}+0.1V_k.
\end{equation}
This formulation balances intra-cluster compactness, inter-cluster separation, and statistical stability, ensuring robust cluster estimation under varying RF channel conditions. 
Clusters with purity above $\tau_p$ and sizes within the range $[s_{\min},s_{\max}]$ are retained as candidate novel UAV categories for the subsequent incremental learning stage. 
The detailed clustering procedure is summarized in Algorithm~\ref{alg:cluster_unknowns}.

\begin{algorithm}[t]
\caption{Clustering for Unknown RF Samples}
\label{alg:cluster_unknowns}
\SetKwInOut{Input}{Input}\SetKwInOut{Output}{Output}
\Input{Embeddings $\mathbf{Z}\!\in\!\mathbb{R}^{N\times d}$ of rejected samples; candidate cluster range $k\in\{0,1,\dots,k_{\max}\}$.}
\Output{Chosen cluster number $k^\star$ and assigned labels $\ell$.}

\textbf{Preprocess:} Standardize $\mathbf{Z}$; \If{$\dim(\mathbf{Z})>64$}{apply PCA to $m=\min(64,d)$ dimensions.\;}

\For{$k \gets 0$ \KwTo $k_{\max}$}{
  Fit K-Means$(k)$ and GMM$(k)$.\;
  Compute composite score
  $Q(k) \gets 0.4\,S_k + 0.3\,\dfrac{CH_k}{1000} + 0.2\,\dfrac{1}{1+DB_k} + 0.1\,V_k$.\;
  Record K-Means inertia: $\mathsf{Inertia}(k)$.\;
}

Identify elbow point $k_{\mathrm{elbow}} \gets \textsc{Elbow}(\mathsf{Inertia})$.\;
Identify best score $k_{\mathrm{score}} \gets \arg\max_k Q(k)$.\;

\uIf{$Q(k_{\mathrm{elbow}}) \ge 0.9\,Q(k_{\mathrm{score}})$}{
  $k^\star \gets k_{\mathrm{elbow}}$.\;
}\Else{
  $k^\star \gets k_{\mathrm{score}}$.\;
}

Assign labels $\ell$ using GMM$(k^\star)$ (or K-Means$(k^\star)$).\;
Filter clusters by purity $\ge \tau_p$ and size in $[s_{\min},\,s_{\max}]$.\;
Retain accepted clusters as new UAV classes $\mathcal{C}_{\text{new}}$.\;
\end{algorithm}

\subsection{Incremental Learning with Lightweight Replay}
Once new UAV classes are identified through clustering, the model incrementally integrates them while maintaining performance on previously learned classes $\mathcal{C}_{0:(t-1)}$. 
Samples are selected according to per-class quality control and replay budget rules to ensure balanced coverage and prevent over-representation. 
During each incremental update, both new and replayed samples are mixed into balanced mini-batches, and the model parameters are optimized by minimizing $L_{\text{CE}}^{(t)}$. This lightweight replay mechanism mitigates catastrophic forgetting by maintaining old feature representations within a compact memory budget.

The overall incremental learning loop, summarized in Algorithm~\ref{alg:incremental_loop}, integrates three main stages. 
(1) \textit{Open-set recognition:} each incoming RF sample $X$ is encoded into a semantic embedding $z = E_{\theta}(X)$, and classified based on Mahalanobis distance to existing class distributions. Unknown samples are stored as $\mathcal{U}$.  
(2) \textit{Novel class discovery:} when $|\mathcal{U}|$ exceeds a threshold, Algorithm~\ref{alg:cluster_unknowns} is invoked to cluster $\mathcal{U}$ into new categories $\mathcal{C}_{\text{new}}$, guided by the optimal cluster number $k^\star$.  
(3) \textit{Incremental model update:} a mixed training set $\mathcal{C}_{\text{new}} \cup \mathcal{M}$ is assembled with per-class replay caps $(\text{old}_{\max}, \text{new}_{\max})$, followed by to minimizing $L_{\text{CE}}^{(t)}$ and update the class statistics.

This closed-loop process enables the system to continuously expand its recognition capability in open-world RF environments, autonomously assimilating novel UAV categories while maintaining stable accuracy on previously learned ones. 
It effectively balances adaptability and retention, achieving robustness against catastrophic forgetting under limited memory constraints.

\begin{algorithm}[t]
\caption{Incremental Learning of Newly Discovered Classes}
\label{alg:incremental_loop}
\SetAlgoLined
\KwIn{Encoder $E_\theta$; stats $\{(\boldsymbol{\mu}_k,\boldsymbol{\Sigma}_k,\tau_k)\}_{k\in\mathcal{C}_0}$; replay memory $\mathcal{M}$ (capacity $M_{\max}$).}
\KwOut{Updated encoder $E_{\theta'}$; expanded class set $\mathcal{C}_0\leftarrow\mathcal{C}_0\cup\mathcal{C}_{\text{new}}$.}

{1) Open-set recognition}\\
\For{each incoming RF sample $X$}{
  $\mathbf{z}\gets E_\theta(X)$;\\
  compute $d_k(\mathbf{z})$ for all $k\in\mathcal{C}_0$;\\
  $k^\star \gets \arg\min_k d_k(\mathbf{z})$;\\
  \eIf{$d_{k^\star}(\mathbf{z}) < \tau_{k^\star}$}
    {assign class $k^\star$;}
    {store $\mathbf{z}$ in unknown buffer $\mathcal{U}$;}
}
{2) Novel class discovery}\\
\If{$|\mathcal{U}|\ge n_{\min}$}{
  run Algorithm~\ref{alg:cluster_unknowns} on $\mathcal{U}$ to obtain $\mathcal{C}_{\text{new}}$; \\ clear $\mathcal{U}$;
}

{3) Incremental model update}\\
Assemble mixed training set from $\mathcal{C}_{\text{new}}\cup\mathcal{M}$ with per-class caps (old\_max, new\_max);\\
Update $\theta\gets\arg\min L_{\mathrm{IL}}$; 
\\recompute $(\boldsymbol{\mu}_k,\boldsymbol{\Sigma}_k,\tau_k)$; \\
update $\mathcal{M}$\;

\end{algorithm}

\section{Results and Discussion}\label{sec:results}
\subsection{Experimental Setup}
The proposed framework is evaluated on a real-world UAV RF dataset \cite{wv7h-sv64-24} to assess its effectiveness across open-set detection, novel class discovery, and incremental adaptation. 
Experiments are designed to measure the system's ability to (i) correctly classify known UAVs, (ii) detect and cluster previously unseen UAV types, and (iii) preserve past knowledge during sequential learning updates. 
Table~\ref{tab:experiment_setup} summarizes the dataset characteristics, network configuration, and key hyperparameter settings used in all experiments.

\begin{table}[t]
\centering
\caption{Experimental Setup and Training Parameters}
\label{tab:experiment_setup}
\renewcommand{\arraystretch}{1.15}
\setlength{\tabcolsep}{4pt}
\begin{tabular}{p{0.3\linewidth} p{0.6\linewidth}}
\toprule
\textbf{Category} & \textbf{Description} \\
\midrule
\textbf{Dataset} & Real-world UAV RF dataset with $24$ drone types across $915$\,MHz, $2.4$\,GHz, and $5.8$\,GHz bands \cite{wv7h-sv64-24}. \\
\textbf{Known / Unknown Classes} & 18 classes used for training (known); $6$ reserved for evaluation (unknown). \\
\textbf{Signal Representation} & Complex I/Q samples transformed to STFT-based spectrograms, normalized to $[0,1]$ (see Section~\ref{sec:system}). \\
\textbf{Encoder Training} & Optimized using composite feature loss Eq.~(\ref{eq:feature_loss}) with $\eta_1=0.5$, $\eta_2=0.3$, $\eta_3=0.2$. \\
\textbf{Optimizer / Batch Size} & Adam optimizer, learning rate $10^{-4}$, mini-batch size $64$. \\
\textbf{Incremental Learning} & Incremental updates use a replay memory with per-class caps of $old_{\max}=0, 5, 10$ and $new_{\max}=60$; overall memory budget $M_{\max}$. \\
\bottomrule
\end{tabular}
\end{table}
\subsection{Open-Set Recognition Performance}
\begin{figure}[t]
    \centering
    \includegraphics[width=0.95\linewidth]{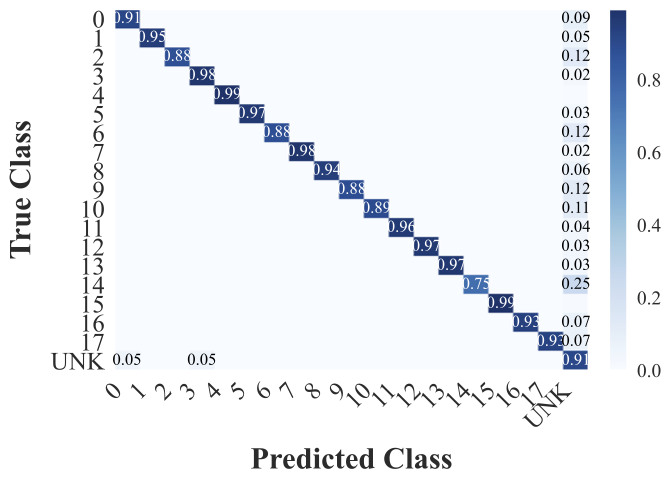}
    \caption{Normalized confusion matrix under open-set testing.}
    \label{fig:confusion_matrix}
\end{figure}

We first evaluate the OSR capability of the proposed framework. Figure~\ref{fig:confusion_matrix} presents the normalized confusion matrix under the open-set condition, where strong diagonal dominance indicates accurate classification of known UAVs. 
Samples from unknown drones are successfully rejected rather than misclassified, confirming that the Mahalanobis-based decision boundary effectively separates unseen categories.

\subsection{Clustering-Based Novel Class Discovery}
\begin{figure*}[h!]
    \centering
    \begin{subfigure}[b]{0.49\textwidth}
        \centering
        \includegraphics[height=4.5cm]{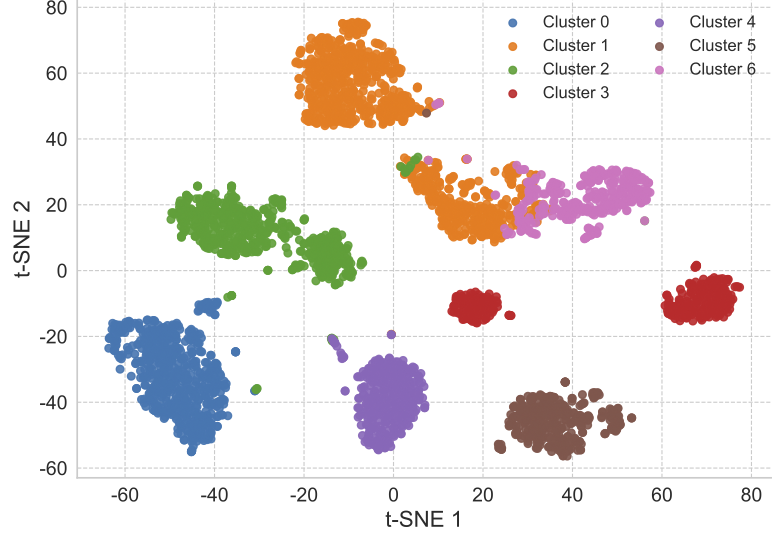}
        \caption{Predicted clusters}
        \label{fig:tsne_predicted}
    \end{subfigure}
    \hfill
    \begin{subfigure}[b]{0.49\textwidth}
        \centering
        \includegraphics[height=4.5cm]{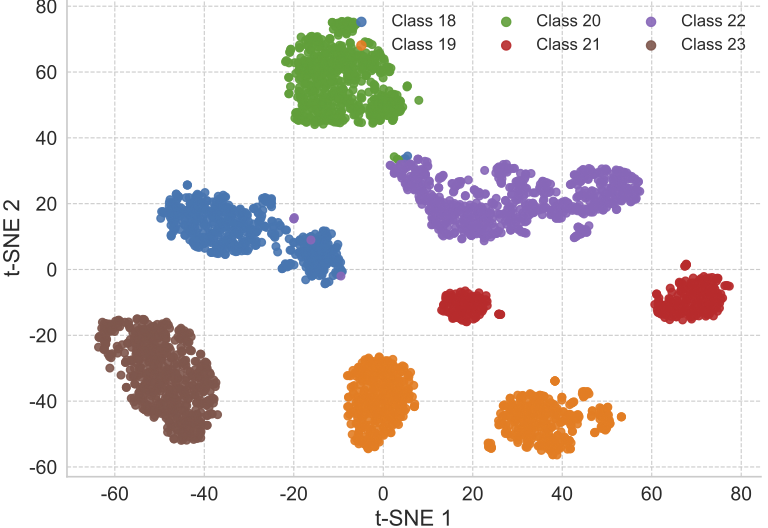}
        \caption{True unknown classes}
        \label{fig:tsne_true}
    \end{subfigure}
    \caption{t-SNE visualization of (a) predicted clusters and (b) true unknown classes. 
    Strong correspondence validates semantic consistency of clustering.}
    \label{fig:tsne_clusters}
\end{figure*}

Figure~\ref{fig:tsne_clusters} visualizes the two-dimensional t-SNE projection of the clustered embeddings from unknown UAV samples. 
Figure~\ref{fig:tsne_clusters}(a) shows the predicted clusters generated by the proposed composite clustering strategy, while Figure~\ref{fig:tsne_clusters}(b) presents the ground-truth unknown UAV classes. 
Each color represents a distinct cluster or true category. 
The strong alignment between Figures~\ref{fig:tsne_clusters}(a) and \ref{fig:tsne_clusters}(b) confirms that the learned feature embeddings preserve semantic relationships among UAV signals, and that the composite cluster-selection strategy effectively partitions unseen UAV types into meaningful groups.
While the overall correspondence between predicted clusters and true unknown classes is strong, a few deviations are observed. 
Specifically, Class~19 appears as the dominant component in both Cluster~4 and Cluster~5, suggesting a mild case of overclustering. 
This can be attributed to the intra-class variability of Class~19, which involves multiple operational modes such as hovering, recording, and navigation that produce distinct RF signatures. 
Conversely, Class~22 is split between Cluster~1 and Cluster~6, where samples from the same UAV type are assigned to two separate clusters. 
This intra-class split likely results from channel or environment dependent variations in the captured signals. 
These observations indicate that although the composite clustering approach maintains overall semantic consistency, signal diversity and propagation effects can still cause mode-dependent fragmentation, suggesting future work on mode-aware or domain-robust clustering mechanisms.

Table~\ref{tab:method-compare} compares the clustering performance of K-Means, GMM, and the proposed composite selection strategy. In this work, the composite mode does not represent a separate clustering algorithm but rather a selection mechanism that evaluates the outcomes of K-Means and GMM using the proposed composite validity score $Q(k)$. This score integrates multiple internal clustering indices as given in \eqref{eq:composite_score}. 
The composite validity score jointly assess intra-cluster compactness, inter-cluster separation, and statistical stability, which is also shown in Algorithm~\ref{alg:cluster_unknowns}.

As shown in Table~\ref{tab:method-compare}, K-Means achieves the highest composite score ($0.770$) and purity ($0.921$), indicating stronger cluster consistency compared to GMM. 
Therefore, under the composite evaluation framework, the K-Means configuration is automatically selected as the optimal clustering result for subsequent incremental learning.

\begin{table}
\centering
\caption{Comparison of clustering modes (K-Means, GMM, Composite).}
\label{tab:method-compare}
\begin{tabular}{lcccccc}
\toprule
Mode & $S_k$ & $CH_k$ & $DB_k$ & $V_k$ & Composite & Purity \\
\midrule
K-Means & 0.444 & 1399.6 & 0.947 & 0.698 & 0.770 & 0.921 \\
GMM & 0.388 & 1222.7 & 1.234 & 0.702 & 0.682 & 0.906 \\
Composite & 0.444 & 1399.6 & 0.947 & 0.698 & 0.770 & 0.921 \\
\bottomrule
\end{tabular}
\end{table}

\subsection{Incremental Learning Performance}

Finally, we evaluate the incremental update stage, where new UAV classes discovered by clustering are assimilated without retraining the entire model. Table~\ref{tab:il} illustrates the effect of replay memory size on old and new class accuracies. When no replay is used ($\texttt{old\_max}=0$), catastrophic forgetting occurs, and old class accuracy drops to 0\%. Introducing a small replay budget ($\texttt{old\_max}=5$) restores the performance to 100\%, while maintaining approximately 98\% accuracy on new classes. Further increasing the replay size yields negligible improvement, indicating that a minimal subset of representative samples suffices to stabilize learning.


\begin{table}
\centering
\caption{Incremental learning accuracy under replay budget constraint.}
\label{tab:il}
\small
\begin{tabular}{@{}lccc@{}}
\toprule
\textbf{Replay cap} & \textbf{Train size} & \textbf{Acc$_\text{old}$ (\%)} & \textbf{Acc$_\text{new}$ (\%)}\\
\midrule
\texttt{old\_max}=0 & $(360,512)$ & $0.0$ & $98.7$\\
\texttt{old\_max}=5 & $(450,512)$ & $100.0$ & $98.1$\\
\texttt{old\_max}=10 & $(540,512)$ & $100.0$ & $98.5$\\
\bottomrule
\end{tabular}
\end{table}

These results confirm that the proposed incremental framework achieves a favorable stability–plasticity trade-off: it retains past knowledge with minimal rehearsal while effectively learning new UAV types. The combination of semantic embedding, probabilistic open-set detection, and selective replay thus provides a robust foundation for real-time UAV classification in evolving airspace environments.
\subsection{Discussion and Limitations}
The current evaluation is conducted on a fixed class split with 18 known and 6 unknown UAV classes to ensure reproducibility. Evaluating multiple known and unknown ratios and random unknown-class selections is an important next step for assessing generalization stability. In addition, the current incremental evaluation considers the introduction of new classes after initial training, while a stricter sequential arrival setting will be explored in future work. For open-set recognition, the Gaussian approximation of class-wise Mahalanobis distances is adopted as a practical thresholding strategy and works well in the evaluated setting, although its validity in more complex RF environments requires further investigation. Finally, the results show that replay is currently the primary mechanism for mitigating forgetting, and future work will explore regularization-based or partially frozen updates to reduce replay dependence.

\section{Conclusion}
This paper presented an incremental open-set learning framework for RF-based UAV recognition that enables reliable operation under dynamic and evolving airspace conditions. 
Unlike conventional closed-set classifiers that fail when unseen drones appear, the proposed system integrates open-set recognition, clustering-based novel class discovery, and replay-driven incremental learning into a unified pipeline. 
The Mahalanobis-distance decision mechanism effectively separates unknown UAV signals, while the composite clustering criterion enables automatic and semantically consistent discovery of new drone categories. 
The lightweight replay strategy further allows the model to assimilate these new classes without retraining from scratch or suffering from catastrophic forgetting. 
Experimental results on a real-world multi-band RF dataset demonstrate that the proposed method achieves high open-set detection accuracy, strong cluster purity, and stable incremental adaptation with minimal memory overhead. 
These findings highlight the potential of incremental open-set learning for scalable and lifelong RF signal intelligence. Future work will explore domain invariant feature extraction under varying channel conditions, real-time deployment on embedded RF receivers, and federated incremental learning across distributed UAV monitoring nodes.

\printbibliography[heading=bibintoc]

\end{document}